\documentclass[aps,prl,twocolumn,letterpaper,superscriptaddress]{revtex4}
\usepackage{graphicx}
\usepackage{epstopdf}
\usepackage{amsmath,color}
\usepackage{amssymb}
\usepackage{textcomp}
\usepackage{url}
\def\be{\begin{equation}}
\def\ee{\end{equation}}
\long\def\symbolfootnote[#1]#2{\begingroup%
\def\thefootnote{\fnsymbol{footnote}}\footnote[#1]{#2}\endgroup}
\DeclareMathOperator{\sgn}{sgn}
\newcommand{\ave}[1]{\langle #1 \rangle}

\def\be{\begin{equation}}
\def\ee{\end{equation}}
\def\bea{\begin{eqnarray}}
\def\eea{\end{eqnarray}}

\begin{document}

\title{
Competing Universalities in Kardar$-$Parisi$-$Zhang Growth Models
}
\author{Abbas Ali Saberi}\email{ab.saberi@ut.ac.ir}
\affiliation{Department of Physics, University of Tehran, P. O. Box 14395-547, Tehran, Iran}
\affiliation{School of Particles and Accelerators, Institute for Research in Fundamental Sciences IPM, Tehran, Iran}
\affiliation{Institut f\"ur Theoretische
  Physik, Universit\"at zu K\"oln, Z\"ulpicher Str. 77, 50937 K\"oln,
  Germany}
\author{Hor Dashti-N.}
\affiliation{School of Physics, Korea Institute for Advanced Study, Seoul 130-722, South Korea.}
\author{Joachim Krug}
\affiliation{Institut f\"ur Biologische Physik, Universit\"at zu K\"oln, Z\"ulpicher Str. 77, 50937 K\"oln,
  Germany}

\date{\today}

\begin{abstract}
We report on the universality of height fluctuations at the crossing
point of two interacting $1+1$-dimensional Kardar-Parisi-Zhang
interfaces with curved and flat initial conditions. We introduce a
control parameter $p$ as the probability for the initially flat
geometry to be chosen 
% during the time evolution of the growth process,
and compute the phase diagram as a function of $p$. We find that
the distribution of the fluctuations converges to the Gaussian
orthogonal ensemble Tracy-Widom (TW) distribution for $p<0.5$, and to
the Gaussian unitary ensemble TW distribution for $p>0.5$. For $p=0.5$
where the two geometries are equally weighted, the behavior is
governed by an emergent Gaussian statistics in the 
% dynamical
universality class of Brownian motion. 
% at long times. 
We propose a phenomenological theory to explain our findings and discuss  
% Our findings open pathways towards theoretical understanding of the
% model and its 
possible applications in nonequilibrium transport and traffic flow.
\end{abstract}

%\pacs{}

\maketitle

Scale invariant fluctuations play a central role in the emergence of
universal properties in complex random systems interconnecting various
areas of physics, mathematics and statistical mechanics. 
% \cite{Krug1992}. 
Whereas the concept of universality classes is well established in
the theory of equilibrium phase transitions \cite{Binder1981}, our
understanding of systems driven out of equilibrium is much less
complete \cite{Henkel2008}. 
The Kardar$-$Parisi$-$Zhang (KPZ) equation \cite{KPZ1986} governing
the evolution of the surface height
$h$(\textbf{x},t), 
\be\label{Eq1}\partial_t h(\textbf{x},t)=\nu \nabla^2
h+\frac{\lambda}{2}(\nabla h)^2+\eta(\textbf{x},t),\ee 
is a prototypical model for describing nonequilibrium growing interfaces with a wide range of theoretical and experimental applications \cite{Krug1991,Stanley1995,HHZ1995,Krug1997}. 
The first term in (\ref{Eq1}) represents relaxation of the interface caused by a
surface tension $\nu$, the second describes the nonlinear growth
locally normal to the surface, and the last term is uncorrelated
Gaussian white noise in space and time with zero average $
\langle\eta(\textbf{x},t)\rangle=0 $ and $\langle
\eta(\textbf{x},t)\eta(\textbf{x}',t')\rangle=2D\delta^d(\textbf{x}-\textbf{x}')\delta(t-t')$,
representing the stochastic nature of the growth process. One recovers the Edwards$-$Wilkinson equation for $\lambda=0$. 

The universality class of randomly growing interfaces is usually
characterized by the scaling exponents defined by Family$-$Vicsek
scaling \cite{Family1985} i.e., $w^2(t,l)\sim t^{2\beta}
f(l/t^{\beta/\alpha})$, in terms of the second moment $w^2(t,l)$ of
the height fluctuations at a measurement scale $l$ at time $t$, where
$f(x)\rightarrow$ const as $x\rightarrow\infty$ and $f(x)\sim
x^{2\alpha}$ as $x\rightarrow 0$. 
Thus $w^2$ grows with time like $t^{2\beta}$ until it saturates to $l^{2\alpha}$ when $t\sim l^{\alpha/\beta}$. The universality class is characterized by the exponents $\alpha$ and $\beta$ (the roughness and the growth exponents, respectively),
whose exact values for the KPZ equation are known only in 1+1
dimensions (1+1 D) as $\alpha=1/2$ and $\beta=1/3$. 

\begin{figure}[]
\includegraphics[width=0.8\columnwidth,clip=true]{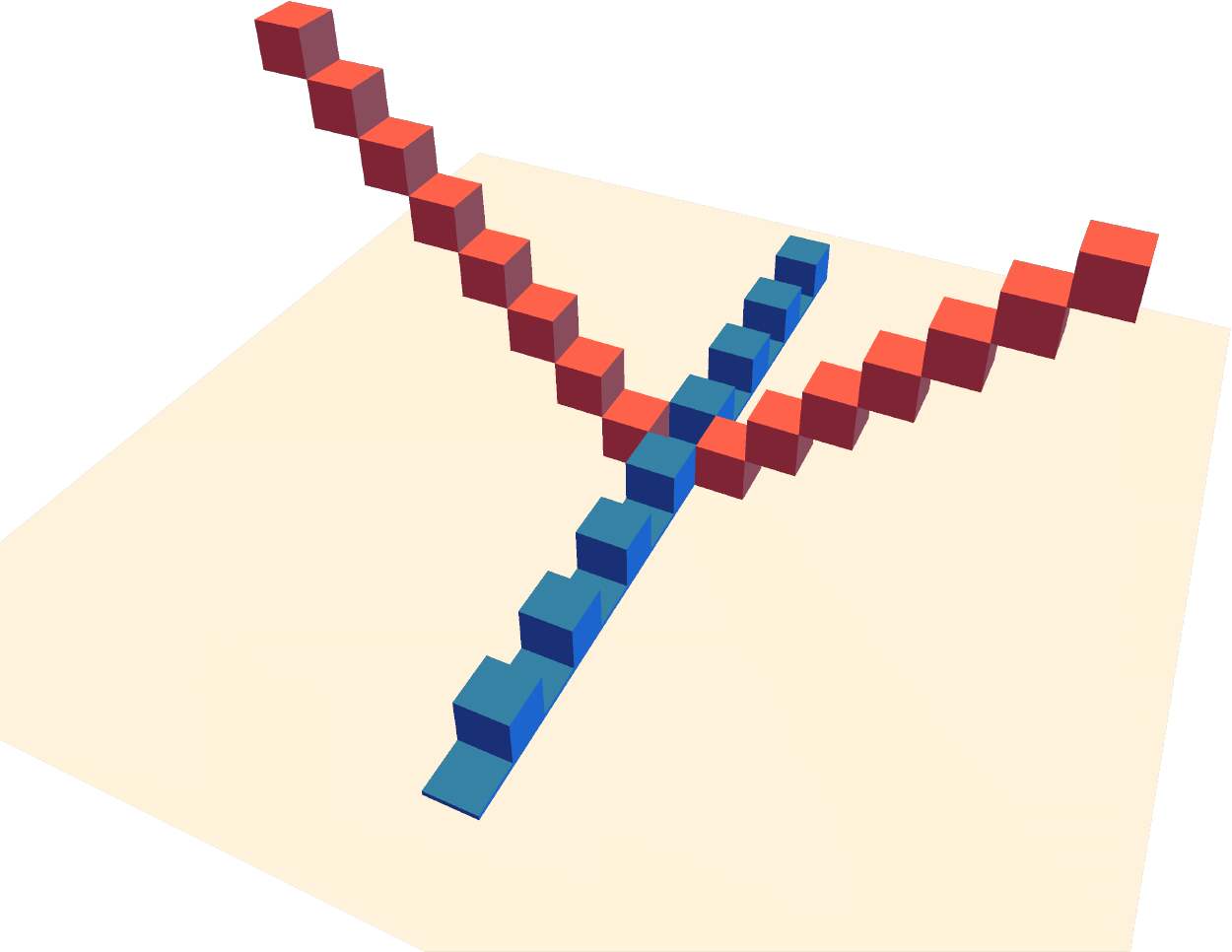}
\caption{\label{fig:fw}
(color online)
 Schematic of the crossing flat$-$wedge geometry with a single common site in the middle.
}\end{figure}

In a series of pioneering works, it has been shown that the universality in various growth models belonging to the KPZ
class holds beyond the second moment
\cite{Krug1992,Krug2010RMT,Takeuchi2011}. Unexpectedly, the height
fluctuations of the 1+1 D single-step
model (SSM) \cite{Meakin1986SSM} grown from a point seed were found to
be governed \cite{Johansson2000} by the Tracy-Widom (TW) distribution of the Gaussian unitary
random matrix ensemble (GUE) \cite{Mehta2004}. Thereafter, it was reported
\cite{prahofer2000statistical,prahofer2000universal} that the radial
1+1 D polynuclear growth (PNG) model also follows the TW GUE
distribution, and in addition, the Gaussian orthogonal ensemble (GOE)
determines the universality of the 1+1 D KPZ growth models on a flat
substrate \cite{prahofer2000universal}. Recently,
exact solutions of the 1+1 D KPZ equation have confirmed the TW GUE
distribution for the height fluctuations on the curved (wedge-like)
\cite{sasamoto2010one,amir2011probability} and the TW GOE distribution on the flat
geometries \cite{calabrese2011exact}. The key question of
interest in this Letter is how these two GOE and GUE universalities
compete when two different 1+1 D KPZ growth models adopting the flat and curved geometries meet each other at a single common point (Fig. \ref{fig:fw}). 

The SSM is a solid on solid
growth model in the KPZ class in which at each time step on a 1 D
(flat or wedge-like) lattice of size $L$, one site $-L/2 \leq j <
L/2 $ is randomly chosen, and if it is a local minimum the height
$h(j)$ is increased by $2$. The initial conditions at $t=0$ are 
% considered as 
$h^f_0(j) = [1-(-1)^j]/2$ and $ h^w_0(j)=|j|$ for the flat and wedge geometries, respectively. This definition guarantees that at each step, the height difference between two
neighboring sites is $\pm 1$. The SSM is the growth model
representation \cite{Rost1981} of the totally asymmetric simple exclusion
process (TASEP) in 1 D, a paradigmatic model for
driven transport of a single conserved quantity \cite{Krug2010RMT}. 
% For a subcase of ASEP known as TASEP (i.e., totally asymmetric simple exclusion process) in which all motion is unidirectional, there has been shown \cite{Johansson2000} that, for step initial condition (similar to the wedge-like initial condition in growth models), the limiting distribution for the current fluctuations converges to the TW GUE distribution. For particles initially positioned every second site (similar to the flat initial condition in growth models), the limit of TASEP is given by the Airy$_1$ process that is stationary with the one-point TW GOE distribution \cite{Borodin2007, Weiss2017}.  

\begin{figure}[t]
\includegraphics[width=0.8\columnwidth,clip=true]{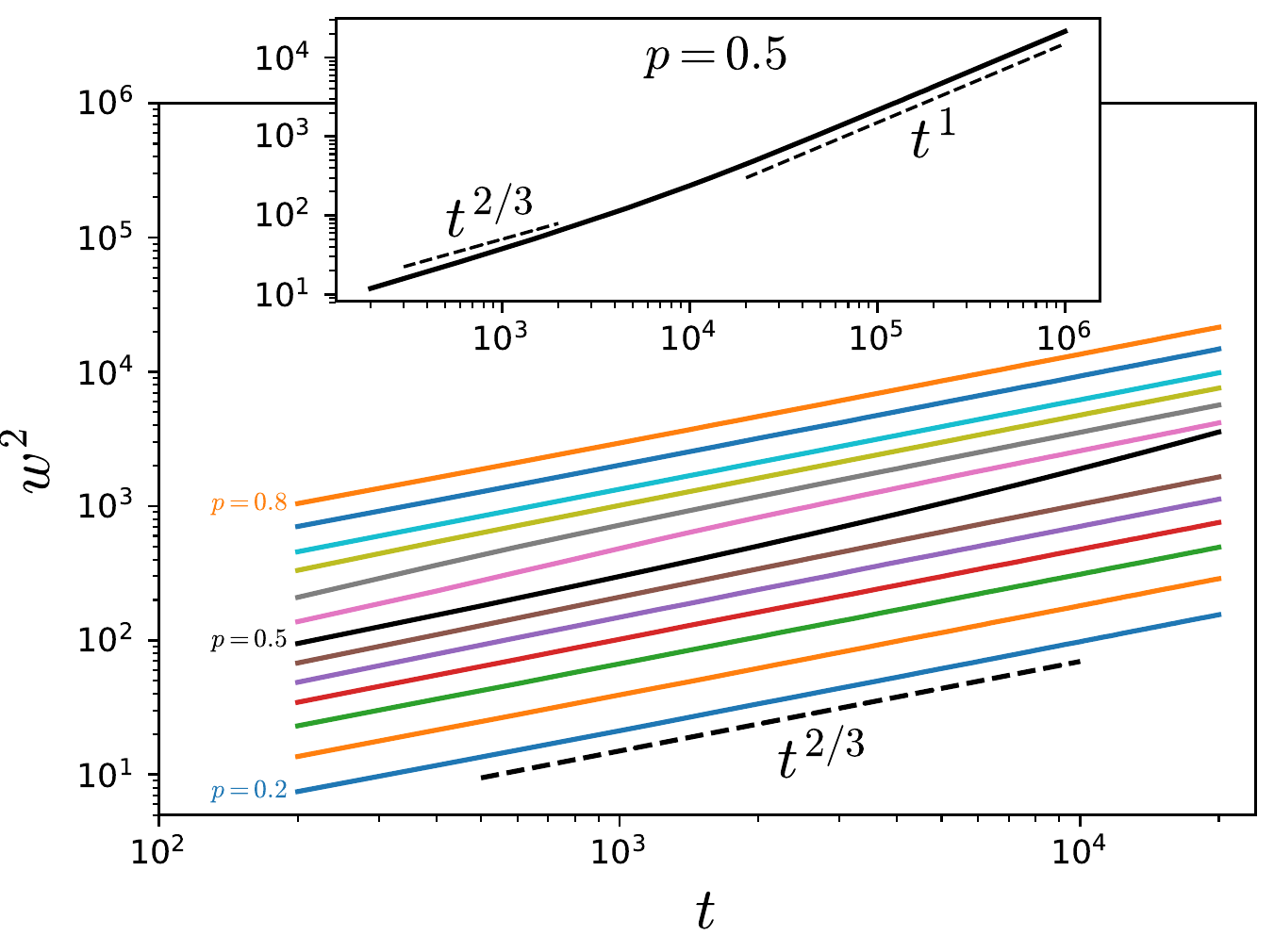}
\caption{\label{fig:fw_w2}
(color online)
 Main: Second moment of the height fluctuations at the crossing point
 of the flat-wedge geometry as a function of time for several $p$ from bottom to top. The
 dashed line shows the scaling prediction $w^2\sim t^{2\beta}$ for the 
1+1 D KPZ equation with growth exponent $\beta=1/3$. All curves are
shifted by a constant for ease of comparison. Inset: The crossover
from 1+1 D KPZ scaling at earlier times to the Brownian motion (BM)
statistics at long time limit for $p=0.5$. In order to clearly observe
the crossover to the BM regime, the simulations for $p=0.5$ were carried out up to time $t=10^6$.  	
}\end{figure}

Here we consider growth on two crossing flat-wedge substrates subject
to the same growth rules but with an exception at the origin
$\textbf{x}=\textbf{0}$, where the two geometries meet. The origin is
the only site with four nearest neighbors, the heights of which have
to exceed the height at $\textbf{0}$ by one for growth to take place. 
% which play role in the growth process of the height there. 
This Letter studies the statistics of the fluctuations of the height
$h(\textbf{0},t)$ at the crossing point at time $t$. Here time is defined
in terms of the number of deposition trials per lattice site, either successful or
not. The initial conditions are set as mentioned above for each
geometry so that $h^f_0(\textbf{0})=h^w_0(\textbf{0})=0$. Periodic
boundary conditions are applied along both geometries. At each time
step, one of the two flat or wedge crossing geometries is chosen with
probability $p$---the only parameter in our study--- and then a site
$j$ is randomly chosen for the growth process.  The flat geometry is
chosen with probability $p$ and the wedge geometry with probability
$1-p$. In the TASEP representation this corresponds to two single-lane
exclusion processes which meet at an intersection. The growth rule at the
origin implies that the particles on the two lanes are forced to cross
the intersection simultaneously.   
% TASEP has also been used in the past as a building block for
TASEP-like traffic flow models with intersections have been studied before,
but with different crossing rules and without considering the current
fluctuations at the intersection \cite{Nagatani1993,Ishibashi1996,Foulaadvand2004,Foulaadvand2007,Belbasi2008,Embley2009,Hilhorst2012,Raguin2013}.

Let us first examine the Family$-$Vicsek scaling for the second moment
of the height fluctuations at the origin i.e., $w^2(t)=\langle
h^2(\textbf{0},t)\rangle-\langle h(\textbf{0},t)\rangle^2$, for
different values of $p$. As Fig. \ref{fig:fw_w2} demonstrates, all
curves for $p\ne0.5$ follow the scaling law $w^2\sim t^{2\beta}$ with
the growth exponent $\beta=1/3$ predicted for the $1+1$ KPZ
equation. A remarkable observation is that for $p=0.5$ when both
geometries are picked with equal probability, the variance of the height at earlier times
behaves as in the KPZ class, but later it crosses over to the
universality of the Brownian motion (BM) i.e., $w^2\sim t$, with
Gaussian statistics (see below).

\begin{figure}[b]
\includegraphics[width=0.8\columnwidth,clip=true]{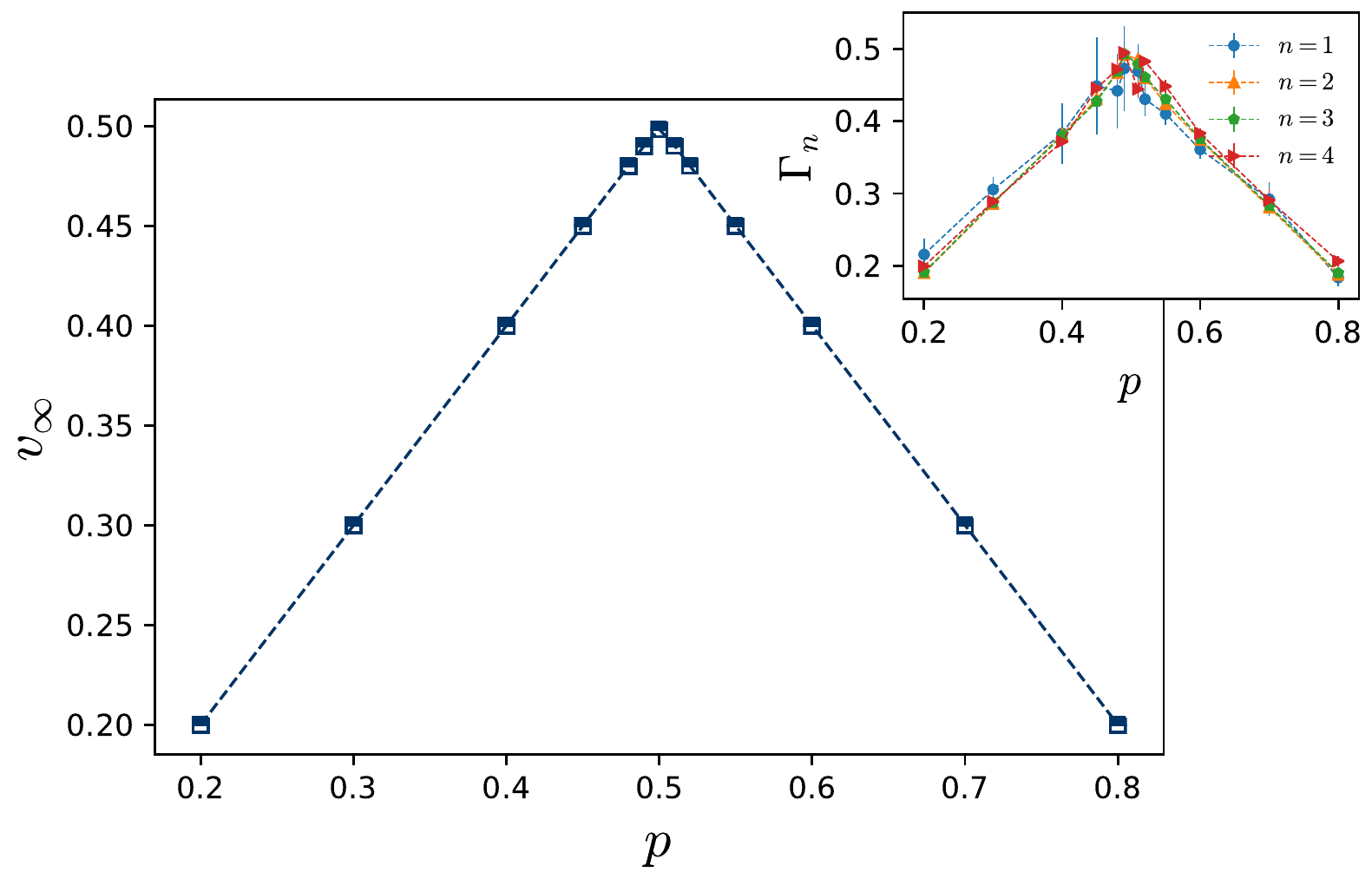}
\caption{\label{fig:v_inf(p)}
(color online)
 $v_\infty$ (main panel) and $\Gamma_n$ (inset) for the flat$-$wedge geometry as a function of $p$.
}\end{figure}

Until now our analysis has revealed two interesting facts: First, the
point with $p=0.5$ acts as a distinguished fixed point with a
characteristic Gaussian statistics in the universality of Brownian
motion, and, second, for $p\ne 0.5$ the statistics of the height
fluctuations at the crossing point---despite the existence of four
nearest neighbors---is compatible with that of the 1+1 D KPZ equation
whose long time statistics converges to the TW GUE/GOE distribution
depending on the narrow-wedge/flat initial condition. One might
naively expect that for $p>0.5$ for which the flat geometry is chosen
with higher probability, the height fluctuations would converge to the
GOE statistics and for $p<0.5$ where the wedge geometry is more likely
to be picked, they should be compatible with the GUE distribution. As we will
show in the following, our results unveil exactly the opposite behavior. 
% Before going further with development of the relation between the random matrix theory and the weighted flat$-$wedge geometry, it is worth mentioning the two special cases i.e., $p=0$ and $p=1$, in which the wedge and flat geometries are always chosen, respectively. In these two cases the height at the crossing point will never evolve in time due to the freezing effect induced by the other non-probable geometry and thus it acts as a pinning point with a fixed small height during the time. This results to a few number of possible height configurations for each case in the order of the system size $\sim\mathcal{O}(L)$. We call these two boundary points as 'locked phases'---Fig. \ref{fig:phasediag}.

The local height of an 1+1 D KPZ interface is asymptotically given
by the following relation \cite{Takeuchi2011},
\begin{equation}\label{eq:h}
h = v_\infty t + s_\lambda (\Gamma t)^{1/3} \chi,
\end{equation}
where $s_\lambda=\sgn(\lambda)$ is the sign of the nonlinear parameter $\lambda$ in the KPZ Eq. (\ref{Eq1}), $ v_\infty $ and $ \Gamma $ are non-universal parameters and $\chi$ is a stochastic variable
with a universal TW distribution depending on the flat/wedge growth
geometry. We estimate the parameter $v_\infty$ by extrapolating
$\ave{h}/t $ versus $t^{-2/3}$, as an intercept in a linear regression
in the $ t \to \infty $ limit, i.e., $\ave{h}/t = v_\infty +
s_\lambda \Gamma^{1/3} \ave{\chi} t^{-2/3}$ \cite{Krug1990}. 
% There exist other approaches to estimate $v_\infty$, but we find this one more conclusive with less fluctuations. 
% In order to have reasonable height ensembles, 
We carried out extensive
simulations to generate height profiles of SSM on the flat$-$wedge
geometry of linear size $L=2^{13}$ up to time $t=2\times 10^{4}$ for
several values $p=0.2$, $0.3$, $0.4$, $0.45$, $0.48$, $0.49$, $0.5$,
$0.51$, $0.52$, $0.55$, $0.6$, $0.7$, $0.8$. For each dataset, an
ensemble of $7\times10^5$ independent realizations have been
generated. 

\begin{figure}[t]
\includegraphics[width=1\columnwidth,clip=true]{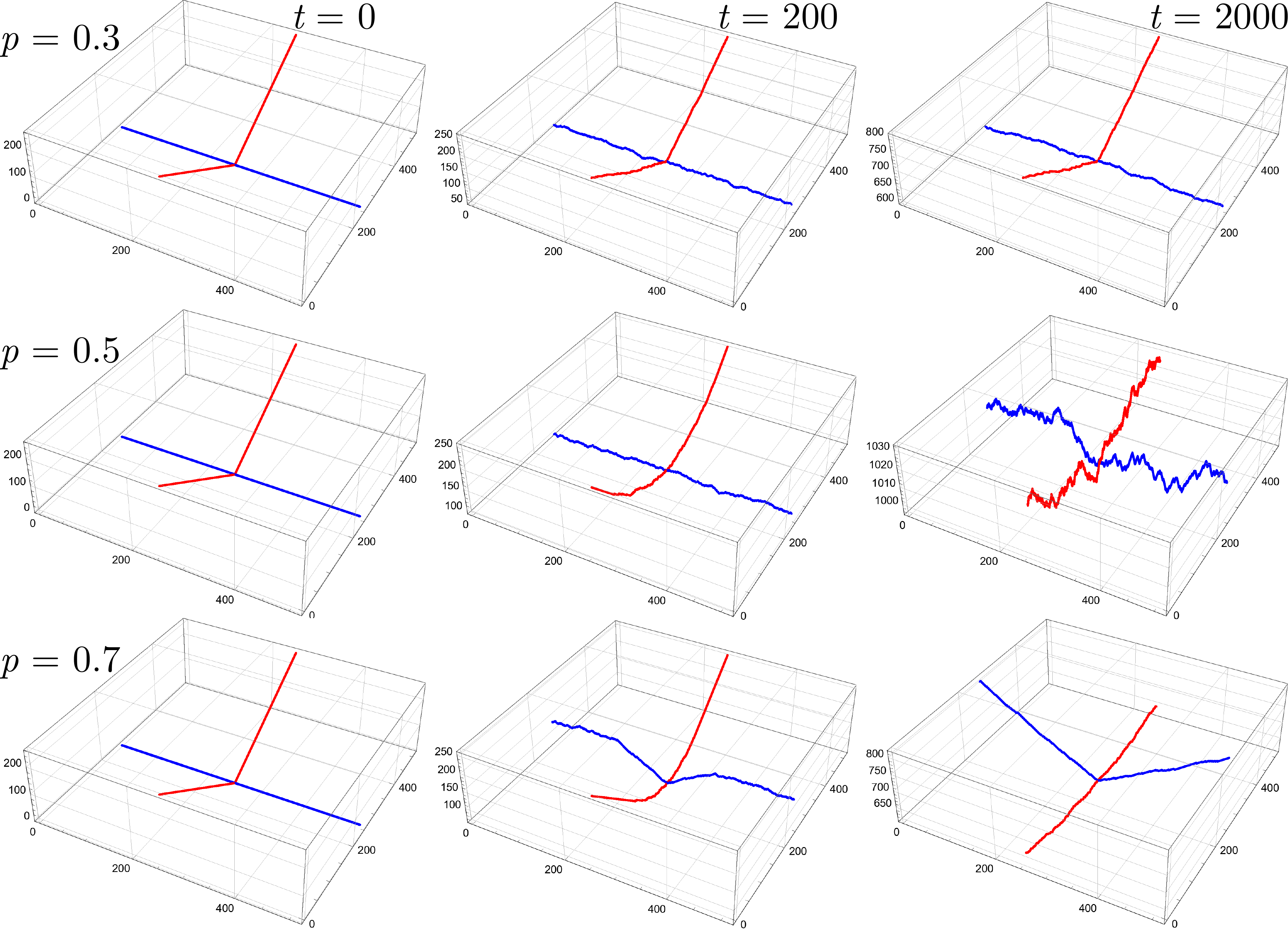}
\caption{\label{fig:ht} 
(color online)
Snapshots for the time evolution of the height profiles on the flat$-$wedge geometry for $t=0$ (left column), $t=200$ (second column), and $t=2000$ (right column) for $p=0.3$ (first row), $p=0.5$ (second row) and $p=0.7$ (third row) corresponding to the GOE, Gaussian (BM) and GUE universality classes, respectively. 
}\end{figure}

As shown in Fig. \ref{fig:v_inf(p)}, we numerically find a
simple relation for $v_\infty$ as a function of the parameter
$p$, 
\begin{equation}
\label{eq:min}
v_\infty(p)=\min(p, 1-p).
\end{equation}
Contrary to the naive expectation, this implies that the substrate
with the \textit{smaller} growth probability dominates the coupled
process. To see why this is so, recall that the asymptotic growth rate
of a single 1+1 D SSM interface with periodic boundary
conditions is given by $v_\infty = \frac{\gamma}{2} (1-u^2)$, where $\gamma$
is the rate of deposition attempts and $u \in [-1,1]$ is the surface
slope \cite{Krug1992,Krug2010RMT}. Because the growth rate is maximal
at $u=0$, an SSM interface can lower its growth rate by developing a
nonzero slope, but it cannot increase its growth rate beyond $\gamma/2$
\cite{Wolf1990,Krug1997}. In the present setting $\gamma = 2p$ for the
flat geometry and $\gamma = 2(1-p)$ for the wedge geometry, respectively. To accomodate a
common growh rate at the origin, for $p < 0.5$ the flat interface
grows at maximal speed $v_\infty = p$ whereas the wedge interface
maintains a nonzero tilt $u = \sqrt{\frac{1-2p}{1-p}}$. For $p > 0.5$ the roles of the two
substrates are interchanged and the initially flat interface becomes
wedge-shaped (Fig.~\ref{fig:ht}).        

\begin{figure}[b]
\includegraphics[width=0.8\columnwidth,clip=true]{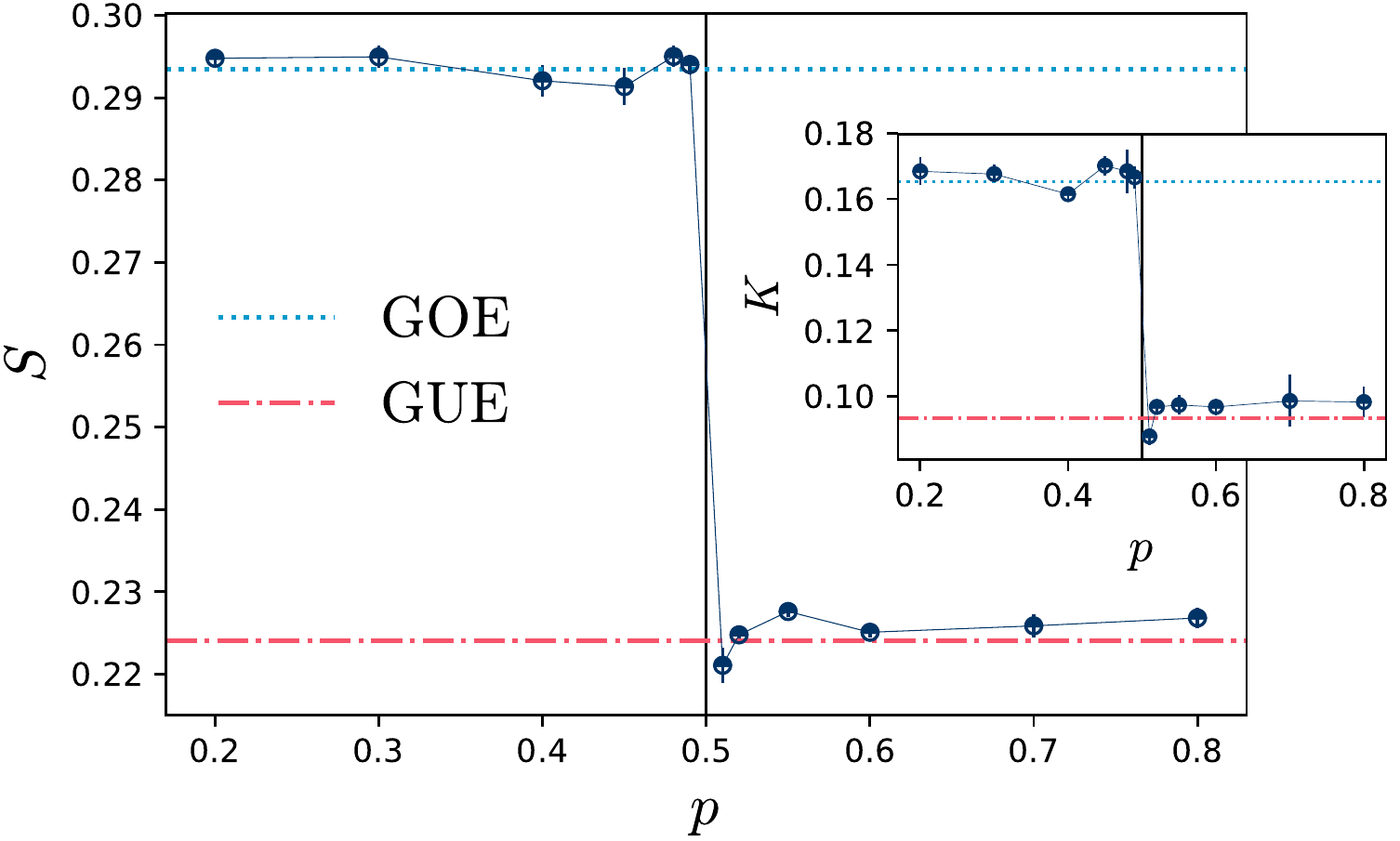}
\caption{\label{fig:fw_SK}
(color online)
 Skewness (main panel) and kurtosis (inset) for the flat$-$wedge geometry as a function of $p$. 
}\end{figure}

We next show that the dominance of the slower geometry extends also to
the height fluctuations at the origin. 
In order to estimate the parameter $\Gamma$ in Eq. (\ref{eq:h}) we define $g_n \equiv \ave{h^n}_c/s^n_\lambda t^{n/3} = \Gamma^{n/3} \ave{\chi^n}_c$, where $ \ave{\chi^n}_c $ denotes the $n$th cumulant of the random variable $\chi$. 
% Alternatively, $\Gamma$ can be obtained from the first cumulant of
% $\chi$, i.e., $g_1 \equiv 3 s_\lambda \left( \ave{h}_t -v_\infty
% \right) t^{2/3} = \Gamma^{1/3} \ave{\chi}$. 
We write $\Gamma_n = [g_n/\ave{\chi^n}_c]^{3/n}$ for the value of
$\Gamma$ estimated from the $n$th cumulant. All estimates have to give
rise to the same value assuming that the cumulants of $\chi$ are those
of the corresponding TW GOE or GUE distributions. 
To find the possible TW distributions, we use two
dimensionless $\Gamma$-independent measures, i.e., the skewness
$S=g_3/g_2^{3/2}$ and the kurtosis $K=g_4/g_2^2$, and compare them
with those of the TW distributions. Figure \ref{fig:fw_SK} represents
the most remarkable finding of our study: For $p<0.5$ the statistics
of the height fluctuations of the crossing point in the wedge$-$flat
geometry is determined by the TW GOE distribution, and, for $p>0.5$ it
is governed by the TW GUE distribution. Therefore we adopt the
corresponding cumulants of the TW distributions into the above
relations to extract $\Gamma_n$. We find that all $\Gamma_n$ follow
the same simple relation with $p$ as we found for $v_\infty(p)$, i.e.,
$\Gamma(p)=\min(p, 1-p)$---see the inset of
Fig. \ref{fig:v_inf(p)}. The relation $\Gamma = v_\infty$ is a known
property of the SSM \cite{Krug1992}. 

\begin{figure}[t]
\includegraphics[width=0.9\columnwidth,clip=true]{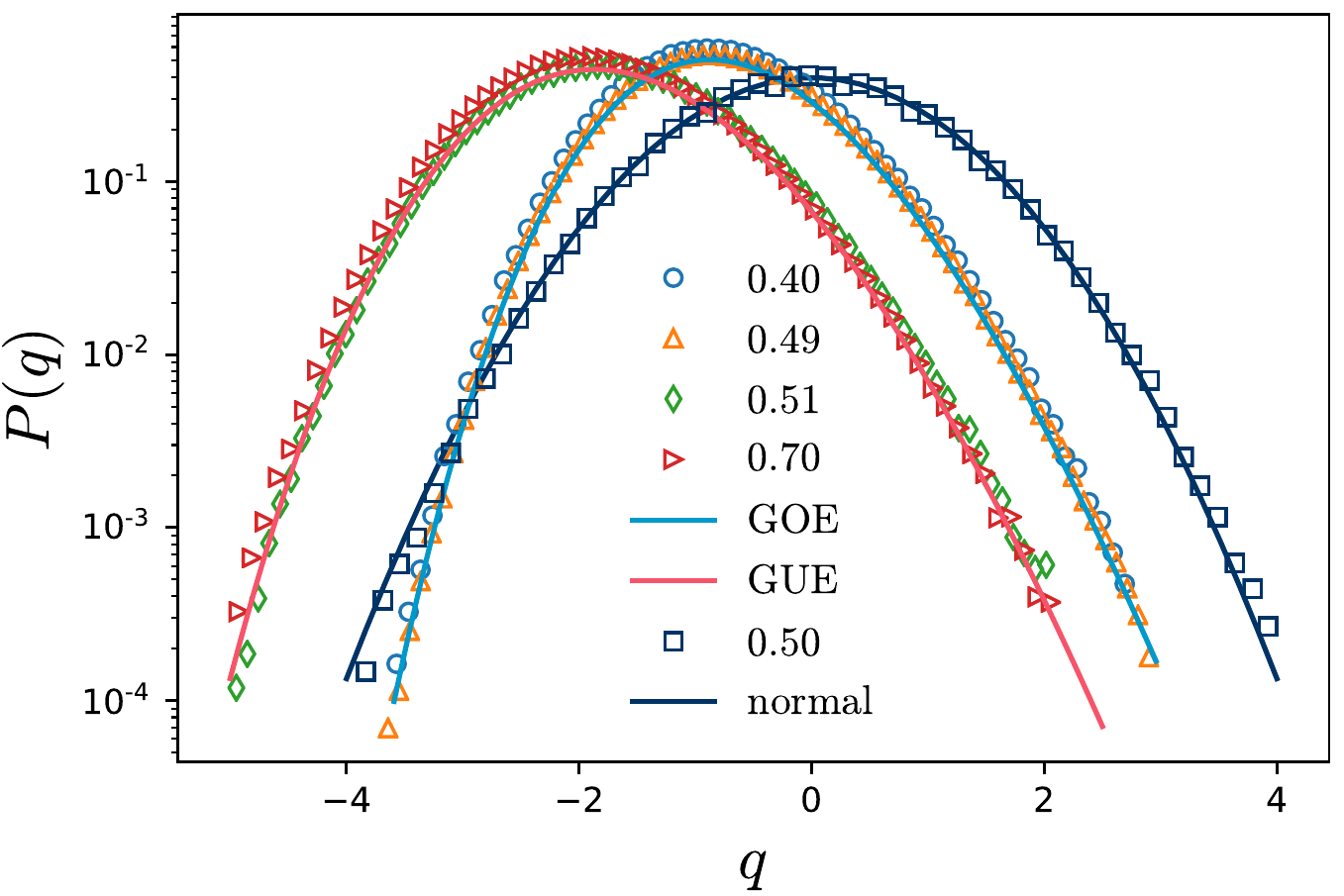}
\caption{\label{fig:fw_P_q}
(color online)
 Rescaled distribution functions of the height fluctuations for the crossing point of the flat$-$wedge geometry for several values of $p$ (symbols), compared with the TW GOE distribution for $p<0.5$, TW GUE distribution for $p>0.5$, and Gaussian distribution for $p=0.5$ (solid lines).
}\end{figure}

Now we can directly check for universality 
% of the height fluctuations
by comparing the height fluctuation distribution with the analytic TW
predictions. For this, we define a new variable $q=(h-v_\infty
t)/s_\lambda (\Gamma t)^{1/3},$ and plot the rescaled distribution
functions $P(q)$ for several values of $p$. Figure \ref{fig:fw_P_q}
shows an excellent agreement with the corresponding TW distributions
for $p\ne 0.5$. The figure also shows the distribution function of
height fluctuations for $p=0.5$ which is in perfect agreement with the
Gaussian distribution. 

The fact that the fluctuations at the crossing point are determined by
the slowly growing interface can be most easily understood in the
TASEP representation. The growth rule at the origin implies that a particle on the fast lane has to
wait for a particle on the slow lane to appear before it can cross the
intersection. Therefore the statistics of the crossing events is
determined by the slower lane, and follows TW-GOE (TW-GUE)
statistics for $p < 0.5$ ($p > 0.5$), respectively. Whereas the
dynamics on the slow lane is asymptotically unaffected by the
intersection, the particles on the fast lane effectively experience a
blockage, which leads to the buildup of a density discontinuity across
the origin. In the interface representation this implies the formation
of a wedge (Fig.~\ref{fig:ht}). 

The physics of inhomogeneous growth processes \cite{Wolf1990} and
exclusion processes with a blockage
\cite{Janowsky1994,Basu2014,Schmidt2015} is also key to understanding
the emergent Gaussian statistics that we observe at $p=0.5$. Consider
first a single, initially flat SSM interface where deposition attempts occur at unit
rate at all sites except a single defect site with deposition rate
$r$. This corresponds to a TASEP with a single slow ($r < 1$) or fast
($r > 1$) bond. Recent work has established that the defect induces a
macroscopic inhomogeneity for any $r < 1$, whereas it is
asymptotically irrelevant when $r > 1$ \cite{Basu2014,Schmidt2015}. We
have numerically studied the height fluctuations at the defect site,
finding TW-GOE statistics for $ r > 1$ but Gaussian BM statistics for
$r < 1$. The latter behavior can be rationalized within the directed
polymer (DP) representation of the process, where the defect site
extends to a defect line in space-time which pins the polymer when 
$r< 1$ \cite{HHZ1995,Krug1997,Basu2014,Tang1993}. In the pinned phase the energy of the polymer,
which translates into the height of the SSM surface, is the sum of
uncorrelated contributions accumulated along the one-dimensional defect
line, which satisfies a central limit theorem and therefore displays
Gaussian statistics. 

The crossing geometry at $p=0.5$ is similar to the SSM with
a defect site, in the sense that deposition occurs at the same rate at
all sites except for the origin, where it is enhanced by a factor of
$r = 2$. By analogy with the 1+1 D SSM, one might anticipate
the existence of a critical value $r_c$, such that the fluctuations
display Gaussian BM statistics for $r < r_c$ and KPZ TW statistics for
$r > r_c$. However, our simulations of a crossing flat-flat geometry
with a variable deposition probability $r$ at the crossing point
indicate that the critical point, which is at $r_c=1$ for the single
lane problem, is shifted to large $r_c\rightarrow\infty$, introducing
the BM statistics as the dominant process in the long-time limit for
any $r$. 
% We believe that 
This may reflect the dynamic nature of the
defect: Even when $r$ is very large, a TASEP particle attempting to
cross the intersection still has to wait for a particle on the second
lane to arrive, which happens at unit rate irrespective of $r$. 
In marked contrast to the 1+1 D SSM, however, we observe 
BM statistics in the absence of a macroscopically tilted,
wedge-like surface profile. To clarify the origin of this behavior, a
DP representation of the crossing growth geometry would be needed.

To conclude, we have considered 1+1 D KPZ growth models on a weighted
flat$-$curved geometry and analyzed the statistics of the height
fluctuations at the crossing point. We found a rich and unexpectedly
non-trivial phase diagram comprising, in addition to the
    known TW GUE/GOE phases, an emergent Gaussian BM phase at $p=\frac{1}{2}$.
% summarized in Fig. \ref{fig:phasediag}. 
It is important to note that the dominance
of the more slowly growing geometry in the SSM is linked to the fact
that the coefficient $\lambda$ of the KPZ nonlinearity is negative in
this case \cite{Krug1992,Wolf1990}. When $\lambda > 0$, the
argument based on the slope-dependence of the asymptotic growth rate
$v_\infty$ predicts that the faster geometry determines the behavior,
which implies that the phase diagram 
% in Fig.~\ref{fig:phasediag} 
is reflected around the point $p=\frac{1}{2}$. We have indeed verified that simulations of the
restricted-solid-on-solid (RSOS) model, which also has $\lambda < 0$, 
lead to the same phase diagram.

At the critical point $p=\frac{1}{2}$, the TASEP representation of the
model relates to previous work on exclusion processes with
intersections \cite{Foulaadvand2007,Embley2009,Raguin2013}, with the seemingly
innocuous modification that particles are forced to cross the
intersection in a correlated manner. Our results suggest that this
makes the transport across the intersections much more efficient, in
that macroscopic density discontinuities do not appear, while a
signature of the intersection is retained in the form of anomalously
large, BM-type current fluctuations. Importantly, the correlated
hopping of particles moving along perpendicular directions is a
fundamental feature of any particle representation of
higher-dimensional growth processes, which is enforced by
the integrability condition on the height field
\cite{Krug1991,Odor2009}. As such, by introducing a single site with a
two-dimensional growth environment into an otherwise one-dimensional
setting, the model may provide an inroad for progress towards
an understanding of the elusive 2+1 D KPZ problem
\cite{HH2012}.
% Our results provide clues for analytical investigations of the model
% that would shed light in better understanding of many other related
% nonequilibrium dynamical processes belonging to the KPZ universality
% class including nonequilibrium transport and  last passage
% percolation models. 

% \begin{figure}[t]
% \includegraphics[width=0.8\columnwidth,clip=true]{fig7.pdf}
% \caption{\label{fig:phasediag}
% (color online)
%  Schematic of the phase diagram for the height fluctuations of the KPZ
% growth models at the crossing point of a flat$-$wedge geometry, where $p$
% denotes the probability for the flat substrate to be chosen.
% At the boundaries $p=0$ and $p=1$ the growth rate at the crossing
% point is strictly zero.
% }\end{figure}

\textit{Acknowledgment.} We thank Andreas Schadschneider for useful
discussions. A.A.S. would like to acknowledge support from the Alexander von Humboldt Foundation and partial financial
support from the research council of the University of
Tehran. J.K. was supported by the German Excellence Initiative through
the UoC Forum \textit{Classical and Quantum Dynamics of Interacting
  Particle Systems.}
\bibliography{refs}

\end{document}